\documentclass{elsartL}
\usepackage{graphicx}
\usepackage{amsmath}
\usepackage{amssymb}
\usepackage{mathrsfs}
\begin{document}


\begin{frontmatter}
\title{On using the Microsoft Kinect$^{\rm TM}$ sensors in the analysis of human motion}
\author[IMS]{M.J.~Malinowski},
\author[IMS]{E.~Matsinos{$^*$}},
\author[InIT]{S.~Roth}
\address[IMS]{Institute of Mechatronic Systems, School of Engineering, Zurich University of Applied Sciences (ZHAW), Technikumstrasse 5, CH-8401 Winterthur, Switzerland}
\address[InIT]{Institute of Applied Information Technology, School of Engineering, Zurich University of Applied Sciences (ZHAW), Steinberggasse 13, CH-8401 Winterthur, Switzerland}

\begin{abstract}The present paper aims at providing the theoretical background required for investigating the use of the Microsoft Kinect$^{\rm TM}$ (`Kinect', for short) sensors (original and upgraded) in the analysis of human 
motion. Our methodology is developed in such a way that its application be easily adaptable to comparative studies of other systems used in capturing human-motion data. Our future plans include the application of this methodology 
to two situations: first, in a comparative study of the performance of the two Kinect sensors; second, in pursuing their validation on the basis of comparisons with a marker-based system (MBS). One important feature in our approach 
is the transformation of the MBS output into Kinect-output format, thus enabling the analysis of the measurements, obtained from different systems, with the same software application, i.e., the one we use in the analysis of 
Kinect-captured data; one example of such a transformation, for one popular marker-placement scheme (`Plug-in Gait'), is detailed. We propose that the similarity of the output, obtained from the different systems, be assessed on 
the basis of the comparison of a number of waveforms, representing the variation within the gait cycle of quantities which are commonly used in the modelling of the human motion. The data acquisition may involve 
commercially-available treadmills and a number of velocity settings: for instance, walking-motion data may be acquired at $5$ km/h, running-motion data at $8$ and $11$ km/h. We recommend that particular attention be called to 
systematic effects associated with the subject's knee and lower leg, as well as to the ability of the Kinect sensors in reliably capturing the details in the asymmetry of the motion for the left and right parts of the human body. 
The previous versions of the study have been withdrawn due to the use of a non-representative database.
\end{abstract}
{$^*$}{E-mail: evangelos[DOT]matsinos[AT]zhaw[DOT]ch, evangelos[DOT]matsinos[AT]sunrise[DOT]ch}
\end{frontmatter}

\section{\label{sec:Introduction}Introduction}

Microsoft Kinect$^{\rm TM}$ (hereafter, simply `Kinect') \cite{Kinect}, a low-cost, portable motion-sensing hardware device, was developed by the Microsoft Corporation (Microsoft, USA) as an accessory to the Xbox $360$ video-game 
console (2010). The sensor is a webcamera-type, add-on peripheral device, enabling the operation of Xbox via gestures and spoken commands. In 2011, Microsoft released the software-development kit (SDK) for Kinect, thus enabling 
the development of applications in several standard programming languages. The first upgrade of the sensor (`Kinect for Windows v2'), both hardware- and software-wise, tailored to the needs of Xbox One, became available for general 
development and use in July 2014. The present paper is part of a broader research programme, aiming at involving either sensor in the analysis of motion data of subjects walking or running `in place' (e.g., on a commercially-available 
treadmill). If successful, Kinect could become an interesting alternative to marker-based systems (MBSs) in capturing data for motion analysis, one with an incontestably high benefit-to-cost ratio.

Regarding medical applications of this sensor, e.g., in physiotherapy in the home environment, a number of products are available, e.g., by `Home Team' (Massachusetts, USA, https://www.hometeamtherapy.com) and `Reflexion' 
(California, USA, http://www.reflexionhealth.com). In particular, it is known that `Reflexion' aims at increasing the success rates in rehabilitation; the participation of the US Navy in the tests of their product attests to 
the importance of the availability of such solutions.

The validation of the output of the original Kinect sensor in static measurements or in case of slow movements was a popular subject in the recent past. Data acquired from $20$ healthy subjects, performing three simple tasks 
(forward reach, lateral reach, and single-leg, eyes-closed standing balance) were analysed in Ref.~\cite{clark}. In their work, the authors drew attention to systematic effects in the Kinect output, in particular for the sternum.

Even more optimistic were the results obtained (and the conclusions drawn) in Ref.~\cite{sysyn}, which investigated the accuracy in the determination of the joint angles from data acquired with an original Kinect sensor and an 
MBS; an inclinometer was assumed to provide the reference or baseline solution (also called `ground truth'). The authors concluded that the differences in accuracy and reliability between the two measurement systems were small, 
thus enabling the use of Kinect as ``a viable tool for calculating lower extremity joint angles''. In fact, the Kinect results and those obtained with the inclinometer were found to agree to better than $2^\circ$. Interestingly, 
the authors also made a point regarding the depth measurements of the Kinect sensor, which are subject to increasing uncertainty with increasing distance from the sensor, reaching a maximal value of about $4$ cm at the most distal 
position of the sensor, which (according to the specifications) should not exceed about $4.5$ m. Naturally, such a dependence introduces bias in the analysis of walking- and running-motion data acquired with a treadmill. These 
effects are present regardless of the largeness of the subject's stray motion in depth; for example, the depth uncertainties are different at the two extreme lower-leg positions, i.e., ahead of and behind the walker/runner.

Another medical application of Kinect was investigated in Ref.~\cite{galna}, namely its use for home-based assessment of movement symptoms in patients with Parkinson's disease. In that study, a number of tasks were performed by 
$19$ subjects, ten of which comprised the control group; parallel data were acquired with Kinect and an MBS. The authors reported that the Kinect results were generally (but not in all cases) found to correlate well with those 
obtained from the MBS for a variety of movements.

Regarding the use of Kinect in medical/health-related applications, complacency and optimism were impaired after the paper of Bonnech${\rm \grave{e}}$re \etal~\cite{bonne} appeared. The authors recorded data from $48$ subjects, 
performing four simple tasks (shoulder abduction, elbow flexion, hip abduction, and knee flexion), and compared the results of different sessions pursuing both their reproducibility within each measurement system, as well as an 
assessment of the differences between the two systems. The authors concluded that the lower body is not tracked well by Kinect. The conclusions of Ref.~\cite{bonne} constitute rather disturbing news in terms of applications of 
the sensor in monitoring walking and running behaviour in a medical/health-related environment.

The literature on the biomechanics of motion is extensive. Some selected works, relevant to the present study, include Refs.~\cite{ck,cl,cbps,o,n,sbbw}. Earlier scientific works are cited therein, in particular in the review 
articles \cite{o,n,sbbw}.
\begin{itemize}
\item Cavagna and Kaneko \cite{ck} studied the efficiency of motion in terms of the mechanical work done by the subject's muscles.
\item Cavanagh and Lafortune \cite{cl} studied the ground reaction forces in running at about $16.2$ km/h, as well as the motion of the `centre of the pressure distribution' during the stance phase of the right foot of $17$ 
subjects. Relatively large variability is seen in their results, partly due to the different characteristics in the motion of rear-foot and mid-foot strikers, partly reflecting the extent of their database in terms of running 
experience, weekly training, and (perhaps, more importantly) habitual individual behaviour. The vertical component of the ground reaction force, which may be as large as three times the subject's body weight, showed sizeable 
variability.
\item Cairns, Burdett, Pisciotta, and Simon \cite{cbps} analysed the motion of ten competitive race-walkers in terms of the ankle flexion, of the knee and hip angles, as well as of the pelvic tilt, obliquity, and rotation. The 
work discussed the main differences between walking and race-walking, and provided explanations for the peculiarity of the motion in the latter case, invoking the goal of achieving higher velocities (than in normal walking) while 
maintaining double support with fully-extended knee and suppressing the vertical undulations of the subject's centre of mass (CM).
\item \~Ounpuu \cite{o} discussed important aspects of the biomechanics of gait, including the variation of relevant physical quantities within the gait cycle. That work may be used as a starting point for those in seek of an 
overview in the topic. It must be borne in mind that the subjects used in Ref.~\cite{o} were children.
\item Novacheck \cite{n} also provided an introduction to the biomechanics of motion. Figs.~5 and 6 of that work contain the variation of the important angles (projections on the coronal, sagittal, and transverse planes) within 
the gait cycle, at three velocities: $4.32$ km/h (walking), $11.52$ km/h (running), and $14.04$ km/h (sprinting). Fig.~9 therein provides the variation of the joint moments and powers (kinesics) in the sagittal plane within 
the gait cycle.
\item In a subsequent article \cite{sbbw}, Schache, Bennell, Blanch, and Wrigley investigated the inter-relations in the movement of the lumbar spine, pelvis, and hips in running, aiming at optimising the rehabilitation process 
in case of relevant injuries.
\end{itemize}

It is rather surprising that only one study, addressing the possibility of involving Kinect in the analysis of walking and running motion (i.e., not in a static mode or in slow motion), has appeared so far \cite{pwbn}. Using 
similar methodology to the one proposed herein, the authors in this study came to the conclusion that the original sensor is unsuitable for applications requiring high precision; after analysing preliminary data, we have come 
to the same conclusion. Of course, it remains to be seen whether any improvement (in the overall quality of the output) can be obtained with the upgraded sensor.

Our aim in the present paper is to develop the theoretical background required for the comparison of the output of two measurement systems used (or intended to be used) in the analysis of human motion; we will give all important 
definitions and outline meaningful tests. Although this methodology has been developed for a direct application in the case of the Kinect sensors, other applications may use this scheme of ideas in order to obtain suitable 
solutions in other cases. The tests we propose in Section \ref{sec:Tests} should be sufficient to identify the important differences in the output of two such measurement systems. As such, they should (in a comparative study) 
pinpoint the essential differences in the performance of the two Kinect sensors or (if the second measurement system is an MBS) enable the validation of the Kinect sensors.

The material in the present paper has been organised as follows. In Section \ref{sec:Systems}, the output of the two Kinect sensors is described; subsequently, the output, obtained with one popular marker-placement scheme from 
an MBS, is detailed. A scheme of association of these two outputs is developed. The definitions of important quantities, used in the description of the motion, are given in Section \ref{sec:Method}. Section \ref{sec:Acquisition} 
describes one possibility for the data acquisition; in the second part of this section, we explain how one may extract characteristic forms (waveforms) from the motion data, representative of the subject's motion within one gait 
cycle. In Section \ref{sec:Tests}, we outline our proposal for the necessary tests, to be performed on the waveforms obtained in the previous section. The last part contains a short summary of the paper and outlines two directions 
in future research.

\section{\label{sec:Systems}The Kinect sensors and the `Plug-in Gait' marker-placement scheme}

To enable the analysis of the data with the same software application, the MBS output, obtained for the specific marker-placement scheme described in Subsection \ref{sec:PiG}, will be transformed into Kinect-output format, using 
reasonable associations between the Kinect nodes and the marker locations; due to the removal of the constant offsets in the data analysis (see Subsection \ref{sec:Analysis3}), the exact matching between the Kinect nodes and the 
locations at which these markers are placed is not essential.

\subsection{\label{sec:Kinect}The Kinect sensors}

\begin{figure}
\begin{center}
\includegraphics [width=15.5cm] {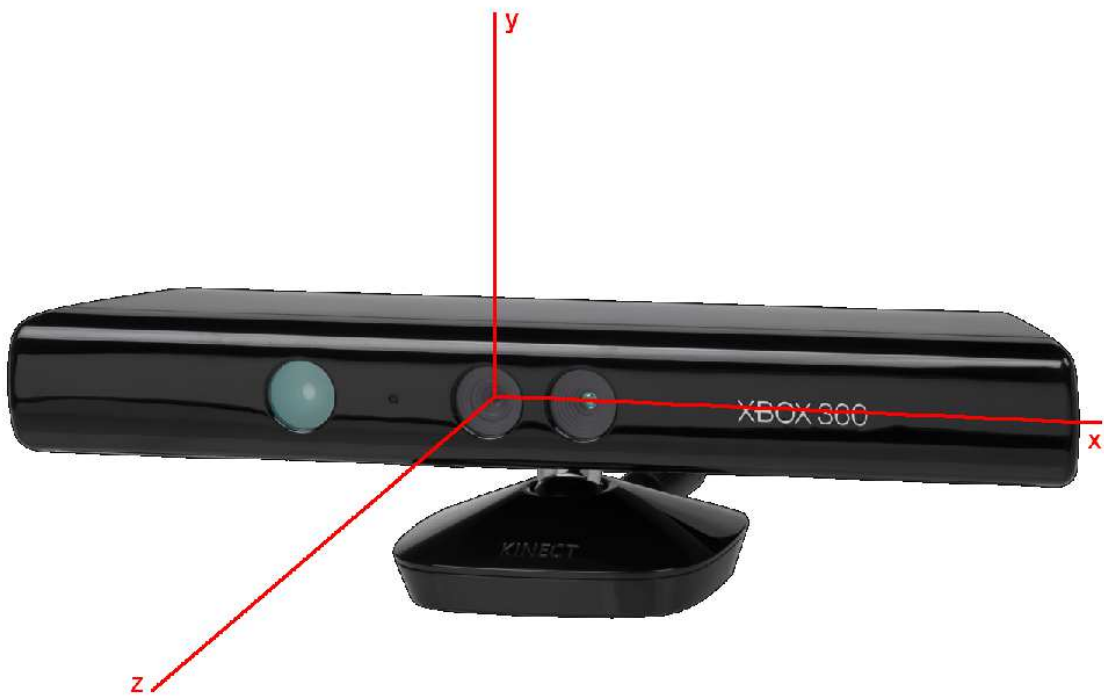}
\caption{\label{fig:Kinect}The front view of the original Kinect sensor; also shown is the Kinect coordinate system.}
\end{center}
\end{figure}

In the original Kinect sensor, the skeletal data (`stick figure') of the output comprises $20$ time series of three-dimensional (3D) vectors of spatial coordinates, i.e., measurements of the ($x$,$y$,$z$) coordinates of the $20$ 
nodes which the sensor associates with the axial and appendicular parts of the human skeleton. In coronal (frontal) view of the subject (sensor view), the Kinect coordinate system is defined with the $x$ axis (medial-lateral) 
pointing to the left (i.e., to the right part of the body of the subject being viewed), the $y$ axis (vertical) upwards, and the $z$ axis (anterior-posterior) away from the sensor, see Fig.~\ref{fig:Kinect}.

\begin{figure}
\begin{center}
\includegraphics [width=15.5cm] {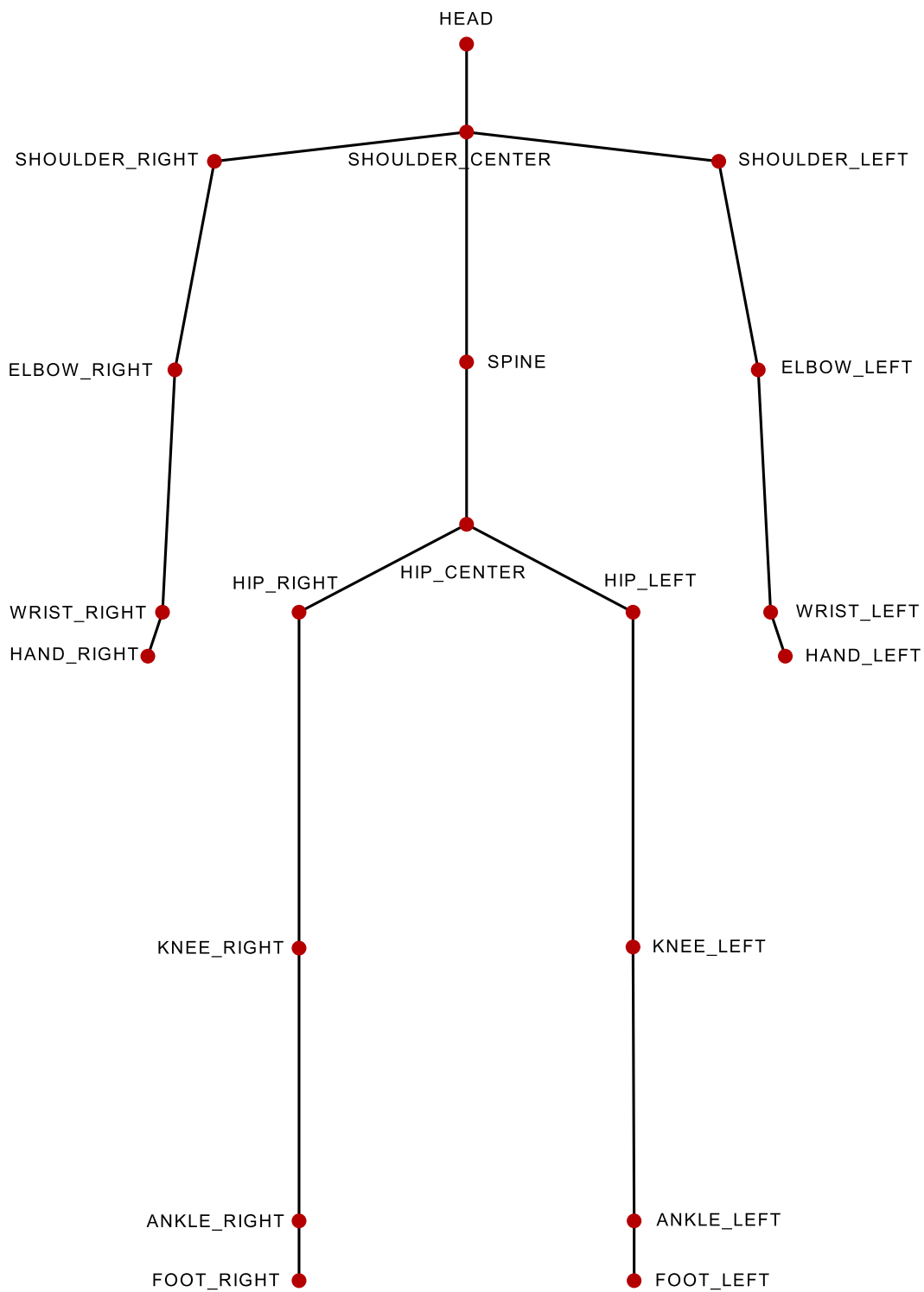}
\caption{\label{fig:NodesKinect}The $20$ nodes of the original Kinect sensor. The figure has been produced with CaRMetal, a dynamic geometry free software (GNU-GPL license), first developed by R.~Grothmann and recently under 
E.~Hakenholz \cite{CMl}.}
\end{center}
\end{figure}

The nodes $1$ to $4$ are main-body nodes, identified as HIP\_CENTER, SPINE, SHOULDER\_CENTER, and HEAD. The nodes $5$ to $8$ relate to the left arm: SHOULDER\_LEFT, ELBOW\_LEFT, WRIST\_LEFT, and HAND\_LEFT; similarly, the nodes 
$9$ to $12$ on the right arm are: SHOULDER\_RIGHT, ELBOW\_RIGHT, WRIST\_RIGHT, and HAND\_RIGHT. The eight remaining nodes pertain to the legs, the first four to the left (HIP\_LEFT, KNEE\_LEFT, ANKLE\_LEFT, and FOOT\_LEFT), the 
remaining four to the right (HIP\_RIGHT, KNEE\_RIGHT, ANKLE\_RIGHT, and FOOT\_RIGHT) leg of the subject~\footnote{The subject's left and right parts refer to what the subject perceives as the left and right parts of his/her body.}. 
The nodes of the original sensor may be seen in Fig.~\ref{fig:NodesKinect}.

In the upgraded Kinect sensor, some modifications have been made in the naming (and placement) of some of the nodes. The original node HIP\_CENTER has been replaced by SPINE\_BASE (and appears slightly shifted downwards); the 
original node SPINE has been replaced by SPINE\_MID (and appears slightly shifted upwards); finally, the original node SHOULDER\_CENTER has been replaced by NECK (and also appears slightly shifted upwards). Five new nodes have 
been appended at the end of the list (which was a good idea, as this action enables easy adaption of the analysis code processing the Kinect output), one of which is a body node (SPINE\_SHOULDER, node $21$), whereas four nodes 
pertain to the subject's hands, HAND\_TIP\_LEFT ($22$), THUMB\_LEFT ($23$), HAND\_TIP\_RIGHT ($24$), and THUMB\_RIGHT ($25$). Evidently, emphasis in the upgraded sensor is placed on the orientation of the subject's hands (i.e., 
on gesturing).

In both versions, parallel to the captured video image, Kinect acquires an infrared image, generated by the infrared emitter (seen on the left of the original sensor in Fig.~\ref{fig:Kinect}); captured with a CCD camera, this 
infrared image provides the means of extracting information on the depth $z$. The sampling rate in the Kinect output (for the video and the skeletal data, for both versions of the sensor) is $30$ Hz.

The description of the algorithm, used in the determination of the 3D positions of the skeletal joints of the subject being viewed by the original sensor, may be found in Ref.~\cite{sh}. Candidate values for the 3D positions 
of each skeletal joint are obtained via the elaborate analysis of each depth image separately. These positions may be used as starting points in an analysis featuring the temporal and kinematic coherence in the subject's motion; 
it is not clear whether such a procedure has been hardcoded in the preprocessing (hardware processing) of the captured data. Shotton \etal~define $31$ body segments covering the human body, some of which are used in order to 
localise skeletal joints, some to fill the gaps or yield predictions for other joints. In the development of their algorithm, Shotton \etal~generated static depth images of humans (of children and adults) in a variety of poses 
(synthetic data). The application of their method results in the extraction of probability-distribution maps for the 3D positions of the skeletal joints; their joint proposals represent the modes (maxima) in these maps. According 
to the authors, the probability-distribution maps are both accurate and stable, even without the imposition of temporal or kinematic constraints. It must be borne in mind that the `3D positions of the joints' of Ref.~\cite{sh} 
are essentially produced from the `3D positions of the projections of the joints onto the front part of the human body' after applying a `shift' in depth (i.e., from the surface to the interior of the human body), namely a 
constant offset ($\zeta_c$) of $39$ mm (see end of Section 3 of Ref.~\cite{sh}). Although the `computational efficiency and robustness' of the procedure are praised in Ref.~\cite{sh}, it remains to be seen whether results of 
similar quality can be obtained in dynamic applications (e.g., when the subject is in motion).

\subsection{\label{sec:PiG}The `Plug-in Gait' marker-placement scheme}

Featuring several cameras, viewing the subject from different directions, MBSs provide powerful object-tracking solutions, yielding high-quality, low-latency data, at frame rates exceeding that of the Kinect sensors. Such 
systems reliably reconstruct the time series of the spatial coordinates of markers (reflective balls, flat markers, active markers, etc.) directly attached to the subject's body or to special attire worn by the subject. One 
popular placement scheme of the markers, known as `Plug-in Gait' \cite{pig}, uses a total of $39$ markers (see Table \ref{tab:MBSMarkers}). The MBS output for these markers may be transformed into Kinect-output format (for 
simplicity, we refer to the naming of the nodes in the original Kinect sensor) by using the following association scheme.
\begin{itemize}
\item The Kinect-equivalent HEAD is assigned to the midpoint of the marker positions LFHD and RFHD. The marker positions LBHD and RBHD, pertaining to the back of the head, are not used.
\item The Kinect-equivalent SHOULDER\_CENTER is taken to be the marker position CLAV. The marker positions C7 and RBAK, which are placed on the back part of the body, are not used in comparisons with data acquired with the original 
Kinect sensor; in the upgraded Kinect sensor, SPINE\_SHOULDER may be identified with C7.
\item The Kinect-equivalent SPINE is estimated as an average of the marker positions T10, LPSI, and RPSI.
\item The Kinect-equivalent SHOULDER\_LEFT and SHOULDER\_RIGHT are taken to be the marker positions LSHO and RSHO, respectively. Regarding the upper part of the body, the marker positions LUPA, LFRA, RUPA, and RFRA are not used.
\item The Kinect-equivalent ELBOW\_LEFT and ELBOW\_RIGHT are taken to be the marker positions LELB and RELB, respectively.
\item The Kinect-equivalent WRIST\_LEFT and WRIST\_RIGHT are assigned to the midpoints of the marker positions LWRA and LWRB, and of RWRA and RWRB, respectively.
\item The Kinect-equivalent HAND\_LEFT and HAND\_RIGHT are taken to be the marker positions LFIN and RFIN, respectively.
\item The Kinect-equivalent KNEE\_LEFT and KNEE\_RIGHT are taken to be the corrected (according to Ref.~\cite{dotg}) marker positions LKNE and RKNE, respectively.
\item The Kinect-equivalent ANKLE\_LEFT and ANKLE\_RIGHT are taken to be the corrected (according to Ref.~\cite{dotg}) marker positions LANK and RANK, respectively.
\item The Kinect-equivalent FOOT\_LEFT and FOOT\_RIGHT are taken to be the marker positions LTOE and RTOE, respectively.
\item The Kinect-equivalent HIP\_LEFT and HIP\_RIGHT positions are evaluated from those of the marker positions LASI, RASI, LPSI, and RPSI, according to Ref.~\cite{dotg}. Regarding the procedure set forth in that paper, a few 
comments are due. The positions of the hips are obtained therein using a model for the geometry of the pelvis, featuring three parameters ($\theta$, $\beta$, and $C$), the values of which had been obtained from a statistical 
analysis of radiographic data of $25$ subjects; however, the values of these parameters are poorly known (see page 583 of Ref.~\cite{dotg}). A simple analysis of the uncertainties given in Ref.~\cite{dotg} shows that, when 
following that method, the resulting uncertainties in the estimation of the positions of the hips are expected to exceed about $10$ mm in each spatial direction. As a result, the positions of the hips, calculated from the MBS 
output according to that procedure, should not be considered as accurate as the rest of the information obtained from the MBS. More importantly, it is not evident how the movement of the pelvis reflects itself in the motion of 
the four markers which are used in the extraction of its position and orientation; it is arguable whether any markers, placed on the surface of the human body, can capture the pelvic motion accurately.
\item The Kinect-equivalent HIP\_CENTER is estimated as an average of the Kinect-equivalent HIP\_LEFT and HIP\_RIGHT, and of the marker position STRN.
\item Regarding the lower part of the body, the marker positions LTHI, LTIB, LHEE, RTHI, RTIB, and RHEE are not used.
\end{itemize}

In regard to the markers placed on the human extremities, it must be borne in mind that their positions are also affected by rotations, not only by the translational motion of these extremities; the markers are placed at some 
distance from the actual rotation axes, coinciding with the longest dimension of the upper- and lower-extremity bones. For instance, rotating the left humerus by $90^\circ$ around its long axis (assumed, for the sake of the 
argument, to align with the vertical axis $y$) will result in a movement of the marker LELB along a circular arc, thus affecting its $x$ and $z$ coordinates. On the other hand, the Kinect nodes are rather placed \emph{on} (or, 
in any case, closer to) the rotation axes; as a result, it is expected that they are less affected by such rotations. As such effects cannot be easily accounted for, it is evident that the association scheme, proposed in the 
present section, can only lead to an approximate comparison of the output of the two measurement systems.

\section{\label{sec:Method}Definitions and scoring options for assessing the similarity of waveforms}

\subsection{\label{sec:Definitions}Definitions of some important angles used in the modelling of the motion}

We will next describe how one may obtain estimates of three important angles in the sagittal plane, representing the level of flexion of the trunk, of the hip, and of the knee. Estimates for the left and right parts of the body 
will be obtained for the hip and knee angles.
\begin{itemize}
\item {\bf Trunk angle}. This angle is obtained from the ($y$,$z$) coordinates of four points, comprising the nodes $1$ (HIP\_CENTER), $3$ (SHOULDER\_CENTER), and two midpoints, namely of the nodes $13$ (HIP\_LEFT) and $17$ 
(HIP\_RIGHT), and of the nodes $5$ (SHOULDER\_LEFT) and $9$ (SHOULDER\_RIGHT). An unweighted least-squares fit on the ($y$,$z$) coordinates of these four points~\footnote{As it is not clear at which depth (and on which basis) 
the original sensor places node $2$ (SPINE), this node should not be included in estimations involving the $z$ coordinate.} yields the slope $\alpha$ (with respect to the $y$ axis) of the optimal straight line. The trunk 
angle is defined as $\theta_T=-\arctan(\alpha)$; $\theta_T=0^\circ$ in the upright position, positive for forward leaning.
\item {\bf Hip angle}. Two definitions of the hip angle have appeared in the literature: the angle may be defined with respect to the trunk or to the $y$ axis; in the present paper, we adopt the latter definition. If the relevant 
hip coordinates are ($y_H$,$z_H$) and those of the knee are ($y_K$,$z_K$), the hip angle is obtained via the expression:
\begin{equation} \label{eq:EQ01}
\theta_H=\arctan \left( \frac{z_H-z_K}{y_H-y_K} \right) \, \, \, .
\end{equation}
Two hip angles will be obtained: the left-hip angle $\theta_{HL}$ uses the nodes $13$ (HIP\_LEFT) and $14$ (KNEE\_LEFT); the right-hip angle $\theta_{HR}$ uses the nodes $17$ (HIP\_RIGHT) and $18$ (KNEE\_RIGHT).
\item {\bf Knee angle}. This is the angle between the femur (thigh) and the tibia (shank). Two definitions of the knee angle have appeared in the literature: the knee angle may be $180^\circ$ or $0^\circ$ in the extended position 
of the knee; we adopt the latter definition. It will shortly become clear why we make use of both the sine and the cosine of the knee angle:
\begin{equation} \label{eq:EQ02}
\beta_1 \equiv \sin(\theta_K)= \frac{(y_A-y_K) (z_K-z_H) - (y_K-y_H) (z_A-z_K)}{L_f L_t}
\end{equation}
and
\begin{equation} \label{eq:EQ03}
\beta_2 \equiv \cos(\theta_K)= \frac{(y_K-y_H) (y_A-y_K) + (z_K-z_H) (z_A-z_K)}{L_f L_t} \, \, \, ,
\end{equation}
where the coordinates of the ankle are denoted as ($y_A$,$z_A$), and $L_f$ and $L_t$ are the projected lengths of the femur and the tibia onto the sagittal plane, respectively:
\begin{equation*}
L_f=\sqrt{(y_K-y_H)^2+(z_K-z_H)^2}
\end{equation*}
and 
\begin{equation*}
L_t=\sqrt{(y_A-y_K)^2+(z_A-z_K)^2} \, \, \, .
\end{equation*}
We define the knee angle as:
\begin{equation} \label{eq:EQ04}
\theta_K = \begin{cases} \arccos(\beta_2), & \mbox{for } \beta_1 > 0 \\ \arcsin(\beta_1) , & \mbox{otherwise} \end{cases} \, \, \, .
\end{equation}
Two knee angles will be obtained: the left-knee angle $\theta_{KL}$ uses the nodes $13$ (HIP\_LEFT), $14$ (KNEE\_LEFT), and $15$ (ANKLE\_LEFT); the right-knee angle $\theta_{KR}$ uses the nodes $17$ (HIP\_RIGHT), $18$ (KNEE\_RIGHT), 
and $19$ (ANKLE\_RIGHT).
\end{itemize}

We define four angles in the coronal plane: the lateral trunk, the lateral hip, the lateral knee, and the lateral pelvic angles; the lateral pelvic angle is also called pelvic obliquity. Estimates for the left and right parts of 
the body will be obtained for the lateral hip and lateral knee angles.
\begin{itemize}
\item {\bf Lateral trunk angle}. The same four points, which had been used in the evaluation of the trunk angle in the sagittal plane, are also used in extracting an estimate of the lateral trunk angle; of course, the ($x$,$y$) 
coordinates of these points must be used now. In addition to these nodes, node $2$ (SPINE) may also be used. The lateral trunk angle is defined with respect to the $y$ axis; $\theta_{lT}=0^\circ$ in the upright position, positive 
for tilting in the positive $x$ direction (tilt of the subject to his/her right).
\item {\bf Lateral hip angle}. This angle describes hip abduction/adduction in the coronal plane. Similarly to the hip angle in the sagittal plane, two definitions of the lateral hip angle are possible: the angle may be defined 
with respect to the trunk or to the $y$ axis; herein, we adopt the latter definition. If the relevant hip coordinates are ($x_H$,$y_H$) and those of the knee are ($x_K$,$y_K$), the lateral hip angle is obtained via the expression:
\begin{equation} \label{eq:EQ05}
\theta_{lH}=-\arctan \left( \frac{x_H-x_K}{y_H-y_K} \right) \, \, \, .
\end{equation}
Two lateral hip angles will be obtained: the lateral left-hip angle $\theta_{lHL}$ uses the nodes $13$ (HIP\_LEFT) and $14$ (KNEE\_LEFT); the lateral right-hip angle $\theta_{lHR}$ uses the nodes $17$ (HIP\_RIGHT) and $18$ 
(KNEE\_RIGHT).
\item {\bf Lateral knee angle}. This is the projection of the angle between the femur and the tibia onto the coronal plane.
\begin{equation} \label{eq:EQ06}
\theta_{lK}= \arcsin \left( \frac{(x_K-x_H) (y_A-y_K) - (x_A-x_K) (y_K-y_H)}{L_f L_t} \right) \, \, \, ,
\end{equation}
where $L_f$ and $L_t$ are now redefined as the projected lengths of the femur and the tibia onto the coronal plane, respectively:
\begin{equation*}
L_f = \sqrt{(x_K-x_H)^2+(y_K-y_H)^2}
\end{equation*}
and
\begin{equation*}
L_t = \sqrt{(x_A-x_K)^2+(y_A-y_K)^2} \, \, \, .
\end{equation*}
The angle is defined positive when, with respect to the femur direction, the ankle appears (in coronal view) `further away' from the subject's body. Of course, two lateral knee angles may be defined, corresponding to the left 
and right parts of the human body, $\theta_{lKL}$ and $\theta_{lKR}$, respectively.
\item {\bf Pelvic obliquity}. This angle is defined as:
\begin{equation} \label{eq:EQ07}
\theta_{lP}=\arctan \left( \frac{y_{HR}-y_{HL}}{x_{HR}-x_{HL}} \right) \, \, \, ,
\end{equation}
where ($x_{HL}$,$y_{HL}$) and ($x_{HR}$,$y_{HR}$) are the ($x$,$y$) coordinates of the left and right hips, respectively.
\end{itemize}

In regard to motion analysis, a few additional angles may be found in the literature: the pelvic tilt and the angle describing the plantarflexion/dorsiflexion of the foot are defined in the sagittal plane; the hip, pelvic, and 
foot rotations in the transverse plane. We do not believe that the Kinect output can yield reliable (if any) information on these quantities. The knee angle, obtained from the 3D vectors ($x_K-x_H$,$y_K-y_H$,$z_K-z_H$) and 
($x_A-x_K$,$y_A-y_K$,$z_A-z_K$), will be called `knee angle in 3D'; it is easily evaluated using expressions analogous to Eqs.~(\ref{eq:EQ02})-(\ref{eq:EQ04}). In view of the fact that the angle between 3D vectors is invariant 
under rotations (SO(3) rotation group) and translations in 3D, the knee angle in 3D is independent of the details regarding the alignment between the relevant coordinate systems (e.g., between the Kinect sensor and the MBS 
coordinate systems).

Two last comments are due.
\begin{enumerate}
\item The trunk angle $\theta_T$ is positive in walking and running; it is difficult to maintain balance if one leans backwards while moving forwards. However, the trunk angle, obtained from the Kinect output, is frequently 
negative. This is due to the fact that the nodes of the Kinect output, which enter the evaluation of $\theta_T$, do not represent locations on the spine.
\item Due to the properties of the knee joint, the knee angle is expected to satisfy the condition $\theta_K \geq 0$. In practice, even in the fully-extended position, $\theta_K$ remains (for many subjects) positive; knee 
hyperextension is a deformity. However, owing to the placement of the nodes by Kinect, the knee angle (estimated from the Kinect output) may occasionally come out negative. To examine further such cases, we retain 
Eq.~(\ref{eq:EQ04}) in the evaluation of the knee angle.
\end{enumerate}
One possibility to avoid these effects is to extract robust measures for the selected physical quantities from the data. For instance, one could use the variation of these quantities within the gait cycle or even their range of 
motion (RoM), i.e., the difference between the maximal and minimal values within the gait cycle. As long as an extremity moves as one rigid object, such measures (being differences of two values) are not affected by a constant 
bias which may be present in the data.

\subsection{\label{sec:Comparison}Scoring options when comparing waveforms}

We propose that the similarity of corresponding waveforms (representing the variation of a quantity within the gait cycle, see Subsection \ref{sec:Analysis3}) be judged on the basis of one (or more) of the following 
scoring options: Pearson's correlation coefficient, the Zilliacus error metric, the RMS error metric, Whang's score, and Theil's score. Assuming that a ($0$-centred) waveform from measurement system $1$ (e.g., from one of the 
Kinect sensors) is denoted by $k_i$ and the corresponding ($0$-centred) waveform from measurement system $2$ (e.g., from the MBS) by $q_i$, these five scoring options are defined in Eqs.~(\ref{eq:EQ08})-(\ref{eq:EQ12}) (for 
details on the original works, see Ref.~\cite{mra}); all sums are taken from $i=1$ to $N$, where $N$ stands for the number of bins used in the histograms yielding these waveforms. (In our analyses, we normally use $N=50$.)
\begin{equation} \label{eq:EQ08}
{\rm Pearson\textrm's \,\, correlation \,\, coefficient} \,\, r = \frac{\sum k_i q_i}{\sqrt{\sum k_i^2} \sqrt{\sum q_i^2}}
\end{equation}
\begin{equation} \label{eq:EQ09}
{\rm Zilliacus \,\, error \,\, metric} \,\, d_z = \frac{\sum \lvert k_i - q_i \rvert}{\sum \lvert q_i \rvert}
\end{equation}
\begin{equation} \label{eq:EQ10}
{\rm RMS \,\, error \,\, metric} \,\, d_{\rm rms} = \frac{\sum (k_i - q_i)^2}{\sum q_i^2}
\end{equation}
\begin{equation} \label{eq:EQ11}
{\rm Whang\textrm's \,\, score} \,\, d_w = \frac{\sum \lvert k_i - q_i \rvert}{\sum \lvert q_i \rvert + \sum \lvert k_i \rvert}
\end{equation}
\begin{equation} \label{eq:EQ12}
{\rm Theil\textrm's \,\, score} \,\, d_t = \frac{\sum (k_i - q_i)^2}{\sum q_i^2+\sum k_i^2}
\end{equation}
In case of identical waveforms (from the two measurement systems), $r=1$; all other scores vanish ($d_z=d_{\rm rms}=d_w=d_t=0$).

Evidently, Whang's score is the symmeterised version of the Zilliacus error metric, whereas Theil's score is the symmeterised version of the RMS error metric. Although the differences between the Zilliacus and the RMS error metric 
are generally small (as are those between Whang's and Theil's scores), we make use of all aforementioned scoring options in our research programme.

Other ways for testing the similarity of the output of different measurement systems have been put forth. For instance, some authors favour the use of the `coefficient of multiple correlation' (CMC) \cite{k,gbwm,g,fcc}. Ferrari, 
Cutti, and Cappello \cite{fcc} define the CMC as:
\begin{equation} \label{eq:EQ13}
{\rm CMC} = \left[ 1 - \frac{\sum_{i=1}^{P} \sum_{j=1}^{W} \sum_{k=1}^{N} (w_{ijk} - \bar{w}_{.jk})^2 / (W N (P-1))}{\sum_{i=1}^{P} \sum_{j=1}^{W} \sum_{k=1}^{N} (w_{ijk} - \bar{w}_{.j.})^2 / (W (P N - 1))} \right]^{1/2} \, \, \, ,
\end{equation}
where the triple array $w_{ijk}$ contains the entire data, i.e., $P W$ waveforms of dimension $N$ ($N$ depends on the gait cycle in Ref.~\cite{fcc}); $P$ is the number of measurement systems being used in the study (`protocols', 
in the language of Ref.~\cite{fcc}) and $W$ denotes the number of waveforms obtained within each measurement system. The averages $\bar{w}_{.jk}$ and $\bar{w}_{.j.}$ in Eq.~(\ref{eq:EQ13}) are defined as:
\begin{equation} \label{eq:EQ14}
\bar{w}_{.jk} = \frac{1}{P} \sum_{i=1}^{P} w_{ijk} \, \, \, ,
\end{equation}
\begin{equation} \label{eq:EQ15}
\bar{w}_{.j.} = \frac{1}{N} \sum_{k=1}^{N} \bar{w}_{.jk} \, \, \, .
\end{equation}
Unlike Pearson's correlation coefficient, `directional information' for the association between the tested quantities is lost when using the CMC in an analysis. In its first definition \cite{theil}, the CMC was bound between 
$0$ and $1$. However, the quantity CMC, obtained with Eq.~(\ref{eq:EQ13}), is frequently imaginary (the ratio of the triple sums may be larger than $1$); this is due to the use of $\bar{w}_{.j.}$, instead of the grand mean (along 
with the normalisation factor $W (P N - 1)$, instead of $(W P N - 1)$), in the denominator of the expression. Importantly, it is unclear how the obtained CMC values relate to the goodness of the association between the tested 
waveforms. The association scheme of Ref.~\cite{a} is arbitrary; there is no theoretical justification for such an interpretation of the CMC results.

The basic problem in testing the similarity of the waveforms lies with the fact that the established tests in correlation theory enable the acceptance or the rejection of the hypothesis that the observed effects can be accounted 
for by an underlying correlation of `strength' $\rho_0$, where $-1<\rho_0<1$. The test when $\rho_0=0$ involves the transformation:
\begin{equation*}
t=r \sqrt{\frac{N-2}{1-r^2}} \, \, \, .
\end{equation*}
The variable $t$ is expected to follow the $t$-distribution (Student's distribution) with $N-2$ degrees of freedom (DoF). The tests when $\rho_0 \neq 0$ involve Fisher's transformation; the details may be found in standard 
textbooks on Statistics. No tests are possible when $\rho_0=1$, i.e., when attempting to judge the \emph{goodness} of the association between waveforms, if ideally the waveforms should be identical. The only tests which can be 
carried out in such a case are those involving $\rho_0=0$, i.e., investigating the presence of a statistically-significant correlation between the tested waveforms when the null hypothesis for no such effects is assumed to hold. 
In practice, the one-sided tests for $N-2=48$ DoF result in the rejection of the null hypothesis at the significance level of $5 \%$ when $r \gtrsim 0.2353$ and at the significance level of $1 \%$ when $r \gtrsim 0.3281$.

Formal, well-defined (in the mathematical sense) ways to compare waveforms do exist. As a general rule, the application of rigorous tests has the tendency to yield significant discrepancies in many cases, even when a judgment 
based on a visual inspection of the tested quantities is favourable. a) One possibility would be to obtain the uncertainties in the histogram bins and make use of a $\chi^2$ function to assess the goodness of the association. 
The variability of the output across different sensors could also be assessed and this additional uncertainty could be taken into account in the tests. b) Another possibility would be to invoke analysis of variance (ANOVA), 
defining the reduced `within-treatments' variation as
\begin{equation} \label{eq:EQ16}
\tilde{V}_w = \sum_{i=1}^{P} \sum_{j=1}^{W} \sum_{k=1}^{N} (w_{ijk} - \bar{w}_{i.k})^2 / (P N (W-1))
\end{equation}
and the reduced `between-treatments' variation as
\begin{equation} \label{eq:EQ17}
\tilde{V}_b = \sum_{i=1}^{P} \sum_{j=1}^{W} \sum_{k=1}^{N} (\bar{w}_{i.k} - \bar{w}_{..k})^2 / (N (P-1)) \, \, \, .
\end{equation}
Appearing in these expressions are two average waveforms: the average waveform obtained with measurement system $i$:
\begin{equation} \label{eq:EQ18}
\bar{w}_{i.k} = \frac{1}{W} \sum_{j=1}^{W} w_{ijk}
\end{equation}
and the grand-mean waveform:
\begin{equation} \label{eq:EQ19}
\bar{w}_{..k} = \frac{1}{P} \sum_{i=1}^{P} \bar{w}_{i.k} \, \, \, .
\end{equation}
The ratio $F=\tilde{V}_b/\tilde{V}_w$ is expected to follow Fisher's distribution with $N (P-1)$ and $P N (W-1)$ DoF. The resulting p-value enables a decision on the acceptance or rejection of the null hypothesis, i.e., of the 
observed effects being due to statistical fluctuation. c) A third possibility would be to histogram the \emph{difference} of corresponding waveforms obtained from the two measurement systems within the same gait cycle $j$; the 
decision on whether the final waveform is significantly different from $0$ can be made on the basis of a number of tests, including $\chi^2$ tests for the constancy and shape of the result of the histogram. Nevertheless, to retain 
simplicity in the present paper, we have decided to make use in the data analysis of the simple scoring options introduced by Eqs.(\ref{eq:EQ08})-(\ref{eq:EQ12}).

\section{\label{sec:Acquisition}Data acquisition and analysis}

\subsection{\label{sec:General}Experimental set-up}

The data acquisition may involve subjects walking and running on commercially-available treadmills. The placement of the treadmill must be such that the motion of the subjects be neither hindered nor influenced in any way by 
close-by objects. Prior to the data-acquisition sessions, the two measurement systems must be calibrated and the axes of their coordinate systems be aligned (spatial translations are insignificant). The measurement systems must 
then be left untouched throughout the data acquisition.

The original Kinect sensor also provides information on the elevation (pitch) angle at which it is set. During our extensive tests, we discovered that this information is not reliable, at least for the particular device we used 
in our experimentation. To enable the accurate determination of the elevation angle of the Kinect sensor, we set forth a simple procedure. The subject stands (in the upright position, not moving) at a number of positions on the 
treadmill belt, and static measurements (e.g., $5$ s of Kinect data) at these positions are obtained and averaged. The elevation angle of the Kinect sensor may be easily obtained from the slope of the average (over a number of 
Kinect nodes, e.g., of those pertaining to the hips, knees, and ankles) ($y$,$z$) coordinates corresponding to these positions. The output data, obtained from the Kinect sensor, must be corrected (off-line) accordingly, to yield 
the appropriate spatial coordinates in the `untilted' coordinate system. To prevent Kinect from re-adjusting the elevation angle during the data acquisition (which is a problematic feature), we attach its body unto a plastic 
structure mounted on a tripod.

It is worth mentioning that, as we are interested in capturing the motion of the subject's lower legs (i.e., of the ankle and foot nodes), the Kinect sensors must be placed at such a height that the number of lost lower-leg 
signals be kept reasonably small. Our past experience dictates that the Kinect sensor must be placed close to the minimal height recommended by the manufacturer, namely around $2$ ft off the (treadmill-belt) floor. Placing the 
sensor higher (e.g., around the midpoint of the recommended interval, namely at $4$ ft off the treadmill-belt floor) leads to many lost lower-leg signals leg (the ankle and foot nodes are not tracked), as the lower leg is not 
visible by the sensor during a sizeable fraction of the gait cycle, shortly after the toe-off (TO) instant.

The Kinect sensor may lose track of the lower parts of the subject's extremities (wrists, hands, ankles, and feet) for two reasons: either due to the particularity of the motion of the extremity in relation to the position of 
the sensor (e.g., the identification of the elbows, wrists, and hands becomes problematic in some postures, where the viewing angle of the ulnar bone by Kinect is small) or due to the fact that these parts of the human body are 
obstructed (behind the subject) for a fraction of the gait cycle. Assuming that these instances remain rare (e.g., below about $3 \%$ of the available data in each time series, namely one frame in $30$), the missing values may 
be reliably obtained (interpolated) from the well-determined (tracked) data. Although, when normalised to the total number of the available values, the untracked signals usually appear `harmless' as they represent a small fraction 
of the total amount of measurements, particular attention must paid in order to ensure that no node be significantly affected, as in such a case the interpolation might not yield reliable results.

A few velocities may be used in the data acquisition: walking-motion data may be acquired at $5$ km/h; running-motion data at $8$ and $11$ km/h. At each velocity setting, the subject must be given time (e.g., $1$ min) to adjust 
his/her movements comfortably to the velocity of the treadmill belt. To obtain reliable waveforms from the Kinect-captured data, we recommend measurements spanning at least $2$ min at each velocity.

\subsection{\label{sec:Details}Details on the data analysis}

The subject's motion is split into two components: the motion of the subject's CM and the motion of the subject's body parts relative to the CM. Of course, the accurate determination of the coordinates of the subject's physical 
CM from the Kinect or MBS output is not possible. As a result, the obtained CM should rather be considered to be one reference point, moving synchronously with the subject's physical CM. Ideally, these two points are 
related via a simple spatial translation (involving an unknown, yet constant 3D vector) at all times; if this condition is fulfilled, the obtained CM may be safely identified as the subject's physical CM, because a constant spatial 
separation between these two points does not affect the evaluation of the important quantities used in the modelling of the motion. At all time frames, we obtain the coordinates of the subject's CM from seven nodes, namely from 
the first three main-body nodes $1$ to $3$, from the shoulder nodes $5$ and $9$, as well as from the hip nodes $13$ and $17$. Being subject to considerable movement in walking and running motion, the node $4$ (HEAD) is not included 
in the determination of the coordinates of the subject's CM. Prior to further processing, the CM offsets ($x_{CM}$,$y_{CM}$,$z_{CM}$) are removed from the data; thus, the motion is defined relative to the subject's CM at all times. 
(The angles, defined in Subsection \ref{sec:Definitions}, involve differences of corresponding coordinates; as a result, they are not affected by the removal of the CM offsets from the data.) The largeness of the `stray' motion of 
the subject may be assessed on the basis of the root-mean-square (rms) of the $x_{CM}$, $y_{CM}$, and $z_{CM}$ distributions.

To investigate the stability of the motion over time, the data may be split into segments. In our data analysis, the duration of these segments may be chosen at will; up to the present time, we have made use of $10$ and $12$ s 
segments in the analysis of the Kinect-captured data. Within each of these segments, information which may be considered `instantaneous' is obtained, thus enabling an examination of the `stability' of the subject's motion at the 
specific velocity (see Subsection \ref{sec:Analysis2}). The symmetry of the motion for the left and right parts of the human body may be investigated by comparing the corresponding waveforms. Finally, the largeness of the motion 
of the extremities may be examined on the basis of the RoMs obtained from these waveforms. We subsequently address some of these issues in somewhat more detail.

\subsubsection{\label{sec:Analysis1}Determination of the period of the gait cycle}

Ideally, the period of the gait cycle $T$ is defined as the time lapse between successive time instants corresponding to identical postures of the human body (position and direction of motion of the human-body parts with respect 
to the CM). (Of course, the application of `identicalness' in living organisms is illusional; no two postures can ever be expected to be identical in the formal sense.) We define the period of the gait cycle as the time lapse 
between successive most distal positions $z$ of the same lower leg (i.e., of the ankle or of the ankle-foot midpoint). The arrays of time instants, at which the left or right lower leg is at its most distal position with respect 
to the subject's instantaneous CM, may be used in timing the waveforms corresponding to the left or right part of the human body.

The period of the gait cycle is related to two other quantities which are used in the analysis of motion data.
\begin{itemize}
\item The stride length $L$ is the product of the velocity $v$ and the period of the gait cycle: $L=v T$.
\item The cadence $C$ is defined as the number of steps per unit time; one commonly-used unit is the number of steps per min. It has been argued (e.g., by Daniels \cite{dan}) that the minimal cadence in running motion should be 
(optimally) $180$ steps per min, implying a maximal period of the gait cycle of $2/3$ s.
\end{itemize}

\subsubsection{\label{sec:Analysis2}Assessment of the stability of the motion}

To examine the constancy of the period of the gait cycle throughout each session (according to our definition, each session involves \emph{one} velocity), the values of the instantaneous period of the gait cycle are submitted to 
further analysis. The overall constancy is judged on the basis of a simple $\chi^2$ test, assessing the goodness of the representation of the input data by one overall average value; the resulting p-value is obtained from the 
minimal value $\chi^2_{\rm min}$ for the given number of DoF, i.e., for the number of data segments reduced by one unit. To assess statistical significance, we favour the use of the p-value threshold of $1 \%$, which is a popular 
choice among statisticians.

\subsubsection{\label{sec:Analysis3}Determination of the waveforms}

Using the time-instant arrays from the analysis of the left and right lower-leg signals (as described in Subsection \ref{sec:Analysis1}), each time series (pertaining to a specific node and spatial direction) is split into 
one-period segments, which are subsequently superimposed and averaged, to yield a representative movement for the node and spatial direction over the gait cycle. Finally, one average waveform for each node and spatial direction 
is obtained, representative of the motion at the particular velocity. The investigation of the asymmetry in the motion rests on the comparison of the waveforms obtained for corresponding left and right nodes, and spatial directions.

Average waveforms for all nodes and spatial directions, representing the variation of the motion of that node (in 3D) within the gait cycle, are extracted separately for the left and right nodes of the extremities; waveforms are 
also extracted for the important angles introduced in Subsection \ref{sec:Definitions}. As mentioned in Subsection \ref{sec:Analysis1}, the time instant at which the subject's left (right) lower leg is at its most distal position 
(with respect to the subject's CM) marks the start of each gait cycle (as well as the end of the previous one), suitable for the study of the left (right) part of the human body. In case that left/right (L/R) information is not 
available (as, for example, for the trunk angle), the right lower leg may be used in the timing. All waveforms are subsequently $0$-centred. The removal of the average offsets is necessary, given that the two measurement systems 
yield output which cannot be thought of as corresponding to the same anatomical locations. For instance, according to the `Plug-in Gait' placement scheme, the markers for the shoulder are placed on top of the 
acromioclavicular joints; the Kinect nodes SHOULDER\_LEFT and SHOULDER\_RIGHT match better the physical locations of the shoulder joints.

The left and right waveforms yield two new waveforms, identified as the `L/R average' (LRA) and the `right-minus-left difference' (RLD); if emphasis is placed on the extraction of asymmetrical features in the motion from the 
Kinect output, the validation of the RLDs is mandatory.

\section{\label{sec:Tests}Comparison of the waveforms obtained from the two measurement systems}

The comparisons of the waveforms obtained for the nodes of the extremities from the two measurement systems, as well as of those obtained for the important angles defined in Subsection \ref{sec:Definitions}, are sufficient in 
providing an estimate of the degree of the association of the output of the systems under investigation. If one of these systems is an MBS, such a comparison enables decisions on whether reliable information may be obtained from 
the tested Kinect sensor (assumed to be the second system); a common assumption in past studies \cite{clark,galna,bonne,pwbn} is that the inaccuracy of the MBS output is negligible compared to that of the Kinect sensor. (Of 
course, to obtain from the marker positions information on the internal motion, i.e., on the motion of the human skeletal structure, is quite another issue; we are not aware of works addressing this subject in detail.) As already 
mentioned, the theoretical background, developed in the present paper, also applies to a comparative study of the two Kinect sensors, identifying the similarities and the differences in their performance, but (of course) it cannot 
easily enable decisions on which of the two sensors performs better. In summary, irrespective of whether one of the two measurement systems is an MBS or not, the same tests are performed, but the interpretation of the results is 
different. We propose tests as follows:
\begin{itemize}
\item Identification of the node levels of the extremities and spatial directions with the worst association (e.g., with a similarity-index value in the first quartile of the distribution) between the waveforms of the two 
measurement systems.
\item Determination of the similarity of the association between the waveforms pertaining to the upper and lower parts of the human body.
\item Determination of the similarity of the association between the waveforms pertaining to the three spatial directions $x$, $y$, and $z$.
\item Determination of the similarity of the association between the waveforms obtained from the raw lower-leg signals.
\end{itemize}
We propose separate tests for the LRA and RLD waveforms (see end of Subsection \ref{sec:Analysis3}); if the reliable extraction of the asymmetry of the motion is not required in a study, one may use only the LRA waveforms. After 
studying the goodness of the association between the waveforms at fixed velocity, velocity-dependent effects may be investigated. We will now provide additional details on each of these tests.

The goodness of the association between the waveforms, obtained from the two measurement systems for the eight node levels of the extremities (SHOULDER, ELBOW, WRIST, HAND, HIP, KNEE, ANKLE, and FOOT) and spatial directions, may 
be assessed as follows. Separately for each of the five scoring options of Subsection \ref{sec:Comparison}, for each velocity setting, and for each spatial direction, the node levels may be ranked according to the goodness of 
the association of the waveforms of the two measurement systems. The node level with the worst association may be given the mark of $0$, whereas the one with the best association the mark of $7$. The sum of the ranking scores 
over all velocities and scoring options yields an $8 \times 3$ `matrix of goodness of the association' ($8$ node levels of the extremities, $3$ spatial directions); entries in this matrix are restricted between $0$ and 
$7 \cdot 5 \cdot N_v=35 \cdot N_v$, where $N_v$ is the number of the velocities used in the data acquisition; further analysis of the entries of this matrix yields relative information on the goodness of the association for the 
node levels of the extremities and spatial directions, e.g., it identifies those pertaining to the first quartile of the similarity-index distribution (worst association).

To assess the similarity of the waveforms of the two measurement systems, obtained for the nodes of the upper and lower extremities, one-factor ANOVA tests may be performed, separately for each of the five scoring options of 
Subsection \ref{sec:Comparison}, on the scores obtained at each velocity setting, for all upper-extremity nodes and spatial directions, and all lower-extremity nodes and spatial directions. The outlined test should be sufficient 
in determining whether the performance between the two measurement systems for the lower part of the human body (in relation to its upper part) deteriorates. It must be also investigated whether the aforementioned results are 
significantly affected after excluding the nodes with the worst association between the waveforms of the two measurement systems.

The goodness of the association between the waveforms, pertaining to the three spatial directions $x$, $y$, and $z$, may be determined after employing ANOVA tests. Similarly to the previous tests, it must be investigated whether 
the results are significantly affected after the exclusion of the nodes with the worst association between the waveforms of the two measurement systems.

\begin{figure}
\begin{center}
\includegraphics [width=15.5cm] {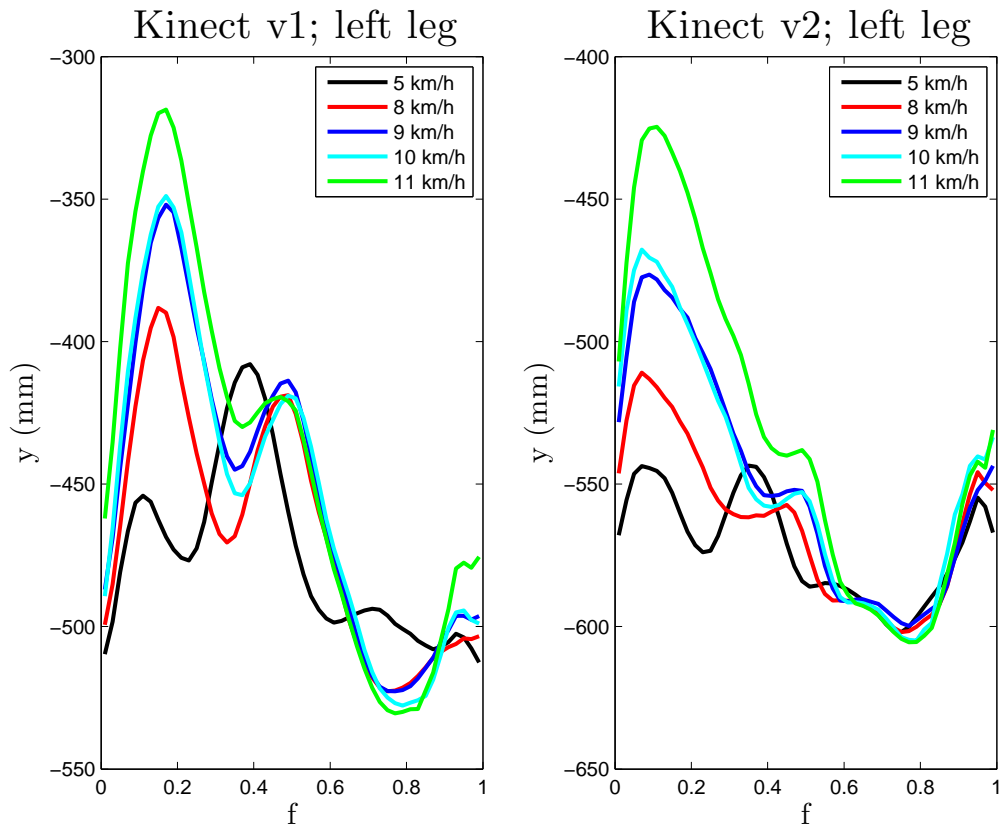}
\caption{\label{fig:LLy_L}Preliminary results for the waveforms for the raw $y$ coordinate of the left lower leg (ankle) obtained from one subject, using both Kinect sensors. The quantity $f$ is the fraction of the gait cycle. 
The sensors were attached unto a plastic structure mounted on a tripod; the difference in the $y$ values simply reflects the higher position on the mount of the upgraded Kinect sensor.}
\end{center}
\end{figure}

\begin{figure}
\begin{center}
\includegraphics [width=15.5cm] {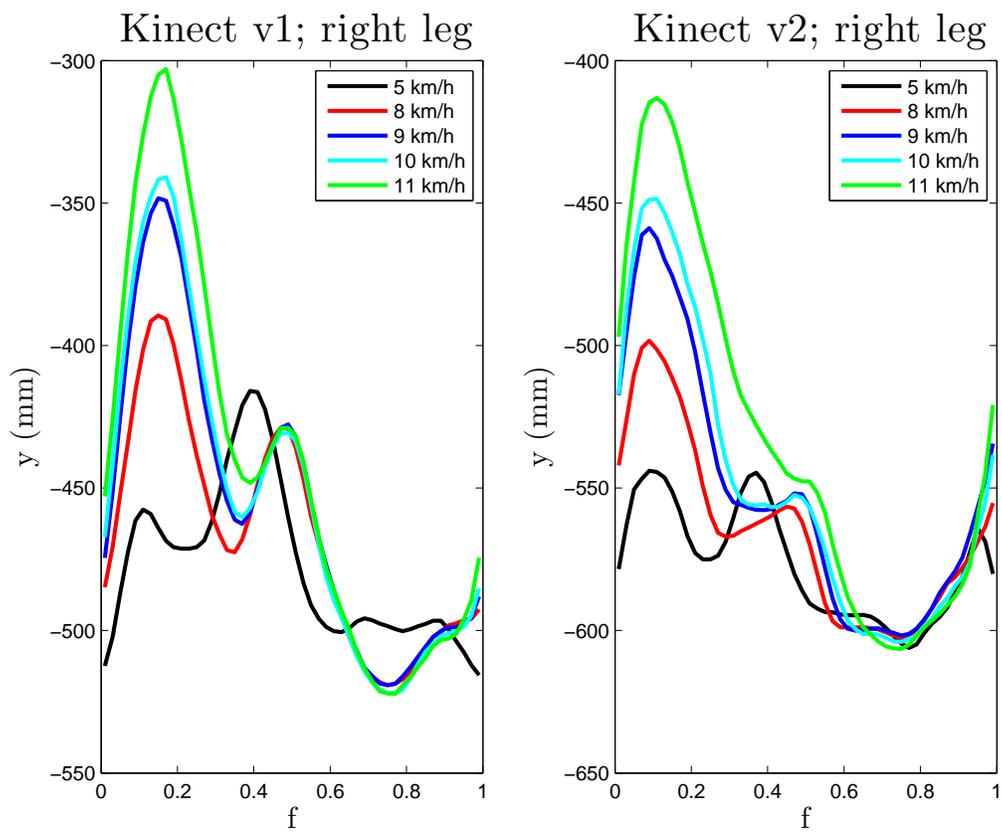}
\caption{\label{fig:LLy_R}Same as Fig.~\ref{fig:LLy_L} for the right lower leg (ankle).}
\end{center}
\end{figure}

Our past experience indicates that the $y$ waveforms, corresponding to the raw lower-leg signals (i.e., the $y$ offsets of the subject's CM are not be removed from the signals), must be examined. This comparison is important for 
two reasons. First, the lower-leg signals are used in timing the motion; second, we intend to use these signals in order to obtain the times (expressed as fractions of the gait cycle) of the initial contact (IC) and the TO 
\cite{o,n}; the difference of these two values is the stance fraction. We had noticed in the past that a salient feature in the waveforms obtained from the original Kinect sensor is a pronounced peak appearing around the IC; this 
peak is less pronounced in the data obtained with the upgraded sensor, e.g., see Figs.~\ref{fig:LLy_L} and \ref{fig:LLy_R}. Although it cannot influence the timing of the motion (because of its position), this artefact complicates 
the determination of the stance fractions, at least when using the original sensor.

The goodness of the association between the RLD waveforms must be investigated in the case that emphasis is placed on the reliable determination of any asymmetric features in the motion. To establish whether the differences in 
the reliability of the LRA and RLD waveforms are significant, two-sided t-tests may be performed on the score distributions between corresponding LRA and RLD waveforms, a total of $15 \cdot N_v$ tests (five scoring options, three 
tests per scoring option, $N_v$ velocity settings). As it is not clear which type of t-tests is more suitable, we propose that three tests be made per case: paired, homoscedastic, and unequal-variance.

Finally, we address the comparison of the RoMs obtained from the waveforms of the two measurement systems. It might be argued that one could simply use in a study the RoMs, rather than the waveforms, as representative of the 
motion of each node. Of course, given that each waveform is essentially replaced by one number, the information content in the RoMs is drastically reduced compared to that contained in the waveforms. Plotted versus one another 
(scatter plot), the ideal relation between the RoMs obtained from the two measurement systems should be linear with a slope equal to $1$, both for the LRA and for the RLD waveforms. The comparison of the two straight-line slopes, 
obtained in case of the LRA and the RLD waveforms, provides an independent assessment on the significance of the differences in the reliability of the LRA and RLD waveforms.

\section{\label{sec:Conclusions}Discussion and conclusions}

Our aim in the present paper was to develop the theoretical background required for the comparison of the output of two measurement systems used (or intended to be used) in the analysis of human motion; important definitions are 
given in Section \ref{sec:Method}, whereas the data acquisition and the first part of the data analysis are covered in Section \ref{sec:Acquisition}. A list of meaningful tests, comprising the second part of the data analysis, is 
given in Section \ref{sec:Tests}.

Although this methodology has been developed for a direct application in the case of the Microsoft Kinect$^{\rm TM}$ (`Kinect', for short) \cite{Kinect} sensors, the use of which in motion analysis is our prime objective, its 
adaption may yield solutions suitable in other cases. The outcome of the proposed tests of Section \ref{sec:Tests} should be sufficient in identifying the important differences in the output of two measurement systems. As such, 
these tests identify (in our case) differences in the performance of the two Kinect sensors (in a comparative study) or enable conclusions regarding the outcome of the validation of the output of either of the Kinect sensors (if 
the second measurement system is a marker-based system (MBS)).

As next steps in our research programme, we first intend to conduct a comparative study of the two Kinect sensors, after applying the methodology set forth herein. At a later stage, we will attempt to validate the output of the 
two Kinect sensors on the basis of standard MBSs.

{\bf Conflict of interest statement}

The authors certify that, regarding the material of the present paper, they have no affiliations with or involvement in any organisation or entity with financial or non-financial interest.

\newpage
\begin{table}[h!]
{\bf \caption{\label{tab:MBSMarkers}}}The notation for the marker positions according to the `Plug-in Gait' placement scheme \cite{pig}.
\vspace{0.6cm}
\begin{center}
\begin{tabular}{|c|c|c|}
\hline
Marker number & Marker-position identifier & Placement \\
\hline
$1$ & LFHD & left front head \\
$2$ & RFHD & right front head \\
$3$ & LBHD & left back head \\
$4$ & RBHD & right back head \\
$5$ & C7 & $7^{\rm th}$ cervical vertebrae \\
$6$ & T10 & $10^{\rm th}$ thoracic vertebrae \\
$7$ & CLAV & clavicle \\
$8$ & STRN & sternum \\
$9$ & RBAK & right back (middle of the right scapula) \\
$10$ & LSHO & left shoulder \\
$11$ & LUPA & left upper arm \\
$12$ & LELB & left elbow \\
$13$ & LFRA & left forearm \\
$14$ & LWRA & left wrist A \\
$15$ & LWRB & left wrist B \\
$16$ & LFIN & left fingers (second metacarpal head, dorsum) \\
$17$ & RSHO & right shoulder \\
$18$ & RUPA & right upper arm \\
$19$ & RELB & right elbow \\
$20$ & RFRA & right forearm \\
$21$ & RWRA & right wrist A \\
$22$ & RWRB & right wrist B \\
$23$ & RFIN & right fingers (second metacarpal head, dorsum) \\
$24$ & LASI & left anterior superior iliac spine \\
$25$ & RASI & right anterior superior iliac spine \\
$26$ & LPSI & left posterior superior iliac spine \\
$27$ & RPSI & right posterior superior iliac spine \\
$28$ & LTHI & left thigh \\
\hline
\end{tabular}
\end{center}
\end{table}

\newpage
\begin{table*}
{\bf Table \ref{tab:MBSMarkers} continued}
\vspace{0.2cm}
\begin{center}
\begin{tabular}{|c|c|c|}
\hline
Marker number & Marker-position identifier & Placement \\
\hline
$29$ & LKNE & left knee \\
$30$ & LTIB & left tibia \\
$31$ & LANK & left ankle \\
$32$ & LHEE & left heel, on the calcaneus \\
$33$ & LTOE & left toes, second metatarsal head \\
$34$ & RTHI & right thigh \\
$35$ & RKNE & right knee \\
$36$ & RTIB & right tibia \\
$37$ & RANK & right ankle \\
$38$ & RHEE & right heel, on the calcaneus \\
$39$ & RTOE & right toes, second metatarsal head \\
\hline
\end{tabular}
\end{center}
\end{table*}

\clearpage

\clearpage

\clearpage

\clearpage

\end{document}